\documentclass[letterpaper, preprint, paper,11pt]{AAS}	

\usepackage{amsmath,amssymb,amsfonts,bm,mathrsfs}
\usepackage{subfigure}
\usepackage[colorlinks=true, pdfstartview=FitV, linkcolor=black, citecolor= black, urlcolor= black]{hyperref}
\usepackage{overcite}
\usepackage{footnpag}			      	

\newcommand{\trace}[1]{\text{trace}(#1)}
\newcommand{\transpose}{^{\top}}
\newcommand{\reals}{\mathbb{R}}

\newcommand{\Euclidnorm}[1]{\|#1\|}

\usepackage{comment}

\PaperNumber{26-682}

\begin{document}

\title{INFORMATION-AWARE MODEL PREDICTIVE CONTROL FOR SATELLITE INSPECTION\footnote{Approved for public release; distribution is unlimited. Public Affairs approval $\#$AFRL-2026-3008. The views expressed are those of the authors and do not reflect the official guidance or
position of the United States Government, the Department of Defense or of the United States
Air Force.
}}

\author{Sarah E. Clees\thanks{Graduate Student, Department of Mechanical and Aerospace Engineering, University of Florida, Gainesville, FL 32611, s.clees@ufl.edu}, 
Sean Phillips\thanks{Technology Advisor, Air Force Research Laboratory - Space Vehicles Directorate, Kirtland AFB, Albuquerque, NM 87117, sean.phillips.9@spaceforce.mil},
\ and
Christopher Petersen\thanks{Assistant Professor, Department of Mechanical and Aerospace Engineering, University of Florida, Gainesville, FL 32611, c.petersen1@ufl.edu}.
}

\maketitle{}

\begin{abstract}
Autonomous spacecraft inspection requires trajectories that satisfy safety and control constraints while enabling the collection of informative measurements about a target spacecraft. Traditional guidance and control methods typically decouple estimation from control, resulting in trajectories that do not explicitly optimize sensing geometry. This work presents a model predictive control (MPC) framework that incorporates estimation covariance in the control objective using a formulation inspired by dual control and covariance steering. The estimation covariance evolves according to a linear Kalman filter, and the measurement model depends on the relative geometry between the agent spacecraft and the target. By embedding the covariance dynamics within the MPC problem, the resulting trajectories account for measurement quality, actively reduce uncertainty, and improve observability in the estimated features of the target. The problem is formulated using relative motion dynamics via the Hill-Clohessy-Wiltshire equations with constraints on control input, relative distance, and terminal maximum covariance. Numerical simulations demonstrate that the proposed framework generates feasible inspection trajectories that actively reduce estimation covariance of points of interest on a target while satisfying input and safety constraints of the agent. A mesh analysis of initial conditions further illustrates how feasibility and the value function of the trajectory depend on the initial conditions and constraint activity.
\end{abstract}

\section{Introduction} Rendezvous and proximity operations (RPO), such as servicing, refueling, and inspection, involve two or more spacecraft operating within close relative distance on the order of 500 km or less \cite{SafeRPO}. For inspection missions in particular, the agent spacecraft must maneuver to observe features of interest on the target while maintaining safe separation and satisfying control input constraints. Autonomous execution of the maneuvers is increasingly necessary due to communication latency between the agent and ground operators \cite{SatComms} along with the dynamic complexity and safety guarantees required for proximity operations \cite{SafeRPO}. As a result, real-time onboard decision-making algorithms must be information-aware to enable autonomous inspection. This paper aims to address this capability by formulating a Model Predictive Control (MPC) framework that explicitly penalizes estimation covariance in the cost function and enforces a terminal upper bound on estimation covariance in the constraints. The problem addressed takes inspiration from the dual control problem \cite{Feldbaum1960,Feldbaum1963,Mesbah2018}. The first aim of the agent's controller is to reduce an exploration cost on estimation covariance of points of interest on the target (to successfully inspect the target). The second aim is to reduce a performance cost on control (to reduce the control effort required to inspect the target). This formulation ensures that the trajectory taken by an agent spacecraft is a path that improves the quality of the information available for estimating a target spacecraft. The measurement model that propagates estimation covariance along the prediction horizon of the MPC framework depends on the agent's position relative to the target via the measurement covariance. Thus, the estimation of the target is coupled with the control policy and the resulting trajectory of the agent. This method ultimately serves as a way to enable information-aware optimization for autonomous spacecraft RPO.

Numerous trajectory planning approaches have been proposed for autonomous inspection missions. Among these methods include optimizing waypoint sequences and natural motion trajectories (NMTs) for agent satellites to travel along, formulated via Markov decision processes and reinforcement learning \cite{Lei_deepRLInspection,Hibbard_CoordinatedInspection, Aurand_deepRLInspection,Aurand_inspectionRegimesActivePassive} along with nonlinear programming \cite{Mercier_optimalNMC}. Such methods have demonstrated the ability to autonomously generate paths around target satellites. However, these approaches often optimize over control effort and inspection time, additionally satisfying sensing and safety constraints, but they do not explicitly optimize the uncertainty in the information collected. Two challenges arise simultaneously for autonomous inspection, as control algorithms must remain feasible under input and safety constraints, while the trajectory must enable measurements that meaningfully reduce uncertainty about the target. An optimal control policy designed solely to satisfy geometric pointing or tracking objectives does not guarantee informative measurements; the geometry of the trajectory directly influences the quality of the data collected, and a metric is needed to quantify successful information gain. Recent work in space domain awareness has emphasized that sensing geometry plays a fundamental role in system observability and the quality of estimation, motivating the design of agent configurations in multi-agent proximity operations that maximize information available \cite{hays2023scitech}. When these concepts of observability are applied to close-proximity inspection missions, this suggests that trajectory design should likewise account for how the relative geometry between the agent and the target influences the quality of the collected information.

In the literature, several approaches exist to quantify and incorporate information into the control loop for autonomous exploration and inspection. Entropy-based metrics \cite{ShannonInfo} provide a theoretically rigorous measure of uncertainty and have been used in path planning and simultaneous localization and mapping (SLAM) based exploration problems \cite{Thrun_robots, Bourgault_infoBasedAdaptive}. However, evaluating expected information gain often requires propagating probability distributions, leading to nonlinear optimization problems that can be computationally challenging for real-time implementation \cite{Bourgault_infoBasedAdaptive}.  As a result, many stochastic planning and control frameworks quantify information through measures of estimation uncertainty, such as the covariance of estimated states, which can be propagated efficiently within recursive estimation algorithms \cite{Simon_OptimEstim,Mesbah2018}. 

SLAM and factor graph formulations from robotics \cite{SingleAgent_GraphBasedSLAM} have been proposed for autonomous spacecraft missions \cite{GraphBased_SLAM_Mercier, SLAM_RPO_RandFiniteSets, SLAM_RPO_LineBasedMonocular}. SLAM-based methods offer the advantage of handling unknown or uncooperative targets by jointly estimating the agent state and a map of environmental features while planning motion relative to the target. However, the dimension of the estimated states grows with the number of mapped features, leading to increased computational burden and memory requirements. As a result, without approximation techniques such as marginalization\cite{SLAMmarginalization} and sliding window filtering \cite{SLAMslidingwindow}, SLAM formulations may be computationally intractable for real-time onboard implementation for spacecraft systems. In contrast, the inspection problem considered in this work assumes that the points of interest to be estimated are selected \textit{a priori} and that the agent state is known. Under these assumptions, the estimation problem reduces to tracking the uncertainty associated with a fixed set of features on the target, making Kalman filter covariance propagation \cite{Simon_OptimEstim} a natural choice for quantifying estimation performance within the control framework.

To explicitly incorporate estimation performance into the trajectory optimization process, this work adopts an information-aware MPC framework inspired by concepts from covariance steering. In classical covariance steering formulations, the control policy shapes the evolution of the mean and covariance of a stochastic system toward a desired terminal distribution \cite{CS_Theory, CovSteering_DiscrtLinearStochasticSyst,CS_ConstrainedDiscrete_LQG}. Extensions to these methods have considered discrete-time stochastic systems with hard input constraints \cite{CS_InputHardConstrained}, probabilistic chance constraints \cite{CS_StochasticSyst_ChanceConstraints,CS_OptimalRiskAllocation}, and finite-horizon stochastic MPC formulations incorporating covariance constraints \cite{CS_Stochastic_MPC,CS_StochLinearDiscrete_FiniteHorizon}. For spacecraft inspection, the measurements taken should improve the agent's knowledge of the target, so incorporating the estimated covariance into the control framework allows the optimizer to steer the distribution of the target state estimation to a desired set. The key difference between standard covariance steering approaches and the present work is that rather than controlling the mean and covariance of the agent to a desired distribution, the stochastic component that is steered is the estimation covariance associated with the target points of interest. Unlike many covariance steering formulations in which uncertainty is propagated from the controlled system dynamics, the dynamics of the agent considered here are deterministic. The covariance evolution of the target points of interest is driven primarily through the Kalman filter measurement update and therefore depends on sensing performance. Because the measurement uncertainty depends on the relative geometry between the inspection agent and the target, the predicted trajectory directly influences the covariance evolution and information gain. By embedding this covariance propagation within the optimization problem, the proposed framework couples estimation performance with the generated inspection trajectory, enabling autonomous spacecraft inspection.

 The primary contribution of this work is an MPC-based inspection framework in which the estimation covariances of points of interest on a target evolve as an explicit function of the inspection trajectory. The key difference between approaches such as dual control and covariance steering and the approach proposed here is that rather than the stochastic parameters being those of the agent dynamics, they are stochastic parameters of the target. Compared to existing inspection methods that involve waypoints and NMTs, the proposed approach explicitly steers the information metric, the estimation covariance, to a desired distribution. This creates a coupled guidance, control, and estimation optimization problem in which the agent trajectory simultaneously satisfies relative motion constraints and safety while actively shaping the information gained about the target. 
 
 The remainder of this paper is organized as follows. First, common notation used throughout the paper is introduced. Then, the problem is motivated and the system dynamics, measurement model, and estimation framework are formulated. Next, the corresponding MPC formulation is presented. Results are provided to demonstrate the proposed approach for an example inspection mission and a feasibility mesh analysis of varying initial conditions, followed by concluding remarks.

\section{Notation}
Let $\reals$ denote the set of real numbers, while $\reals_+$ specifies the set of positive real numbers, and $\mathbb{W}$ is the set of whole numbers. $b\in\reals$ is a real-valued scalar, and $\vec x\in\reals^n$ is a column vector of $n$ real-valued elements. The hat notation $\hat e$ specifies that a vector is a unit-normalized vector. $\mathbf A\in\reals^{n\times n}$ is an $n\times n$ matrix of real-valued elements, $\mathbf I_a$ is the identity matrix of dimension $a$, and $\mathbf 0_a$ is a square matrix of dimension $a$ containing only zero-valued elements. $\mathbf H\in\reals^{3\times 6}$ is a matrix that extracts the relative position states out of the agent state vector. $|b|$ is the absolute value of $b\in\reals$, $\vec x\transpose$ is the transpose operator applied to a column vector, which produces a row vector, $\|\vec x\|$ is the 2-norm defined as $\sqrt{\vec x\transpose \vec x}$, and $\|\vec x\|_\infty$ is the $\infty$-norm defined as $\max_{j=1,2,\dots,n}|x_j|$ for $\vec x\in\reals^n$, where $j$ is an index of the vector. For a matrix $\mathbf P\in\reals^{q\times q}$, trace$(\mathbf P)$ is the trace, defined as the sum of the diagonal elements of the square matrix. For a vector or matrix defined as $\vec x_i=\vec x(t_i),\mathbf V_i=\mathbf V(t_i)$, the subscript denotes the time index. Specifically, when $i$ is used as a subscript, it denotes time in the real world, and when $k$ is used as a subscript, it denotes time in the optimizer prediction horizon. States in the local vertical local horizontal (LVLH) frame are specified with subscripts $x,y,z$. Within the Kalman filter steps, $\vec\theta^-_i$ is an \textit{a priori} estimate at time index $i$ and $\vec\theta^+_i$ is an \textit{a posteriori} estimate at time index $i$. $\vec v\sim\mathcal{N}(0,\mathbf V)$ is a random-valued vector with a zero-mean Gaussian distribution and covariance $\mathbf V$.  

\section{Problem Statement}
Consider the problem of steering an agent spacecraft to collect information about points of interest located on a target spacecraft. These points can be features such as corners of solar panels or other interesting geometries that contribute to the knowledge of the shape and layout of the target. The relative motion of the agent as viewed from the target follows the Hill-Clohessy-Wiltshire (HCW)\cite{HCW} equations and the estimates of the states of the points of interest are propagated by a Kalman filter\cite{Simon_OptimEstim}. The goal of the proposed approach is to reduce the estimation uncertainty of the target states, characterized by the covariance propagated by the Kalman filter, to a desired distribution using MPC.

Assume the agent and target dynamics are characterized by a discrete-time model that has deterministic, linear time-invariant (LTI) dynamics in the agent state 
\begin{equation}
    \vec x_{i+1} = \mathbf{A}_d\vec x_i + \mathbf{B}_d \vec u_i, \label{discreteDynDef}
\end{equation} stochastic LTI dynamics in the information collected about the target 
\begin{equation}
     \vec \theta_{i+1} = \mathbf A_\theta\vec\theta_i + \vec w_i, \label{targetState}
\end{equation} and stochastic linear dynamics with time-varying measurement covariance 
\begin{equation}
    \vec y_{i} = \mathbf C\vec \theta_i + \vec v_i, \label{targetMeasurement}
\end{equation} where $i\in\mathbb{W}$ denotes the discrete time index for time $t_i$. The deterministic quantities are the relative state $\vec x_i=\vec x(t_i)\in\reals^n$ and control input $\vec u_i=\vec u(t_i)\in\reals^m$. The stochastic quantities include the target state $\vec\theta_i=\vec\theta(t_i)\in\reals^q$ which stores the translational coordinates of the $a$ target points of interest ($q=3a$), the measurement vector $\vec y_i=\vec y(t_i)\in\reals^p$, the process noise $\vec w_i=\vec w(t_i)\in\reals^q,~\vec w_i\sim\mathcal{N}(0,\mathbf W)$, and the measurement noise $\vec v_i=\vec v(t_i)\in\reals^p,~\vec v_i\sim\mathcal{N}(0,\mathbf V(\vec x(t_i))$, whose covariance depends on the relative position of the agent with respect to the target.
    
The matrices $\mathbf A_d\in\reals^{n\times n}$, $\mathbf B_d\in\reals^{n\times m}$ follow from HCW dynamics, $\mathbf A_\theta\in\reals^{q\times q}$ represents the dynamics of the points of interest of the target in the local vertical local horizontal (LVLH) frame, and $\mathbf C\in\reals^{p\times q}$ represents the output matrix, which is the relationship between the measurements and the states of the points of interest. The formulation of the measurement covariance $\mathbf V_i\in\reals^{p\times p}$ is discussed in the measurement model formulation below.

\subsection{Relative Motion Dynamics} 
The dynamics are viewed in the LVLH frame fixed to the target. Let the target's position and velocity in an inertial frame be denoted as $\vec{R}$ and $\vec{V}$, respectively. The LVLH frame is defined by three unit vectors that form a basis
\begin{align}
    \hat{e}_x = \frac{\vec R}{\Euclidnorm{\vec R }},~~
    \hat{e}_z = \frac{\vec R \times \vec V}{\Euclidnorm{\vec R \times \vec V}},~~
    \hat{e}_y = \hat e_z \times \hat e_x,
\end{align} where $x$ is the radial direction, $y$ is the along-track direction, and $z$ is the cross-track direction of the target's orbit \cite{Vallado_Astro}. 

\begin{figure}
    \centering
    \includegraphics[width=0.6\linewidth]{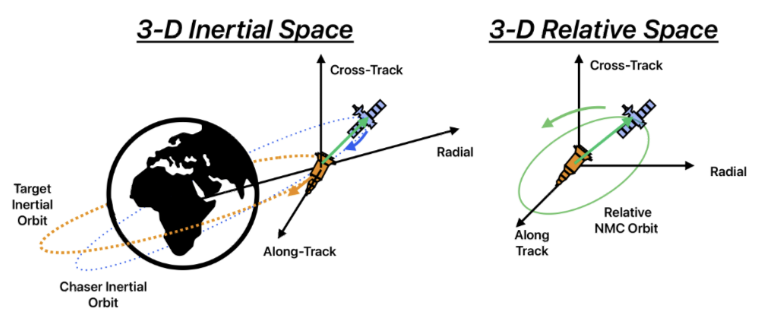}
    \caption{Inertial and LVLH frame representations of motion due to the HCW equations \cite{SafeRPO}. This particular orbit leads to a natural motion circumnavigation (NMC) where the motion viewed in the LVLH frame is closed.}
    \label{fig:NMC}
\end{figure}

 The continuous-time HCW equations for the relative motion of the agent with respect to the target, in the LVLH frame, are defined as a linear system

\begin{equation}
    \dot {\vec x}(t) = \mathbf A_c\vec x(t) + \mathbf B_c\vec u(t) \label{continuousODE},
\end{equation} where the state is
\begin{equation}
    \vec x(t)=\begin{bmatrix}
        r_x & r_y & r_z & \dot r_x & \dot r_y & \dot r_z
    \end{bmatrix}\transpose,
\end{equation}
and the control input is \begin{equation}
    \vec u(t)=\begin{bmatrix}
        F_x & F_y & F_z
    \end{bmatrix}\transpose,
\end{equation} where $\vec r(t)=\left[r_x~~r_y~~r_z\right]\transpose$ defines the relative position vector of the agent and $\dot{\vec r}(t) =\left[\dot r_x~~\dot r_y~~\dot r_z\right]\transpose$ defines the relative velocity vector. The subscripts correspond to the axes of the LVLH frame.
The state and control matrices are defined as
\begin{equation}
    \mathbf A_c = \begin{bmatrix}
        0 & 0 & 0 & 1 & 0 & 0 \\
        0 & 0 & 0 & 0 & 1 & 0 \\
        0 & 0 & 0 & 0 & 0 & 1 \\
        3n^2 & 0 & 0 & 0 & 2n & 0 \\
        0 & 0 & 0 & -2n & 0 & 0 \\
        0 & 0 & -n^2 & 0 & 0 & 0 \\
    \end{bmatrix}, ~~\mathbf B_c= \frac{1}{m}\begin{bmatrix}
        0 & 0 & 0 \\
        0 & 0 & 0 \\
        0 & 0 & 0 \\
        1 & 0 & 0 \\
        0 & 1 & 0 \\
        0 & 0 & 1 \\
    \end{bmatrix},
\end{equation} where $n\in\reals$ is a constant defining the mean motion [rad/s] of the target, assuming that the target is in a circular orbit about Earth (Figure \ref{fig:NMC}), and $m\in\reals$ is the mass of the agent. 
 Assuming the target's orbit and the agent's mass are held constant for the duration of the mission, the dynamics remain time invariant. Assuming that discrete dynamics equations are a valid approximation of the continuous dynamics for short time periods, a discretization of the equations of motion leverages limited computation for propagation of this state space system in time. The solution to Eq. \eqref{continuousODE}, assuming continuous thrust over a time arc $\Delta t$, yields 
\begin{align}
    \mathbf A_d &=e^{\mathbf A_c \Delta t} \label{discreteStateMatrix},\\
    \mathbf B_d &=\int_0^{\Delta t}e^{\mathbf A_c(\Delta t-\tau)}\mathbf B_cd\tau \label{discreteControlMatrix},
\end{align} which will be utilized as the state and control matrices for the discrete system in Eq.~\eqref{discreteDynDef} \cite{LinSystTheory}.

\subsection{Target State and Measurement Model} To develop the measurement model, the following assumptions are made:
\subsubsection{Assumption 1 - Target Shape and Orientation.} 
To simplify the measurement model, assume that the target to inspect is a cube that does not rotate in the LVLH frame. This orientation implies that the state matrix of the target in Eq.~\eqref{targetState} is identity, i.e.
\begin{equation}
    \vec\theta_{i+1}=\vec\theta_i+\vec w_i.
\end{equation}  

\subsubsection{Assumption 2 - Target Points of Interest.} 
Assume that the points of interest of the target are the corners of the cube. This implies that there are eight points of interest, each with a position represented in the LVLH frame, so the number of states to estimate is $q=24$. 

\subsubsection{Assumption 3 - Direct Measurements.} 
The problem is simplified by assuming that noisy measurements taken of the target are the position of the points. This implies that the output matrix of Eq.~\eqref{targetMeasurement} is identity, i.e.
\begin{equation}
    \vec y_i=\vec \theta_i + \vec v_i.
\end{equation}  Thus, $p=q$ in terms of dimensionality. From Assumption 2, the total number of measurements is $p=24$.
\subsubsection{Assumption 4 - Measurement Covariance.}
To represent measurement quality as a function of distance between the agent and target, the measurement covariance is defined as an isotropic function of the agent's relative position. Let the matrix $\mathbf H\in\reals^{3\times 6}$ transform the agent state vector $\vec x_i$ to a vector of the relative positions, i.e.
\begin{equation}
    \mathbf H =\begin{bmatrix}
        \mathbf I_3 & \mathbf 0_3 \\
    \end{bmatrix},~~\mathbf{H}\vec x_i=\begin{bmatrix}
        r_x~r_y~r_z
    \end{bmatrix}\transpose,
\end{equation} where $\mathbf I_3$ is identity and $\mathbf 0_3$ is a square matrix of zeros. Then the measurement covariance is
\begin{equation}
    \mathbf V_i=\mathbf V(\vec x_i) = (\zeta||\mathbf H\vec x_i||^2+\varepsilon)\mathbf I_p, \label{measurementCovDefinition}
\end{equation} where $\mathbf I_p$ is identity, $\zeta\in\reals_+$ is a positive constant scaling factor, and $\varepsilon\in\reals_+$ is the minimum measurement variance. Due to this formulation, the covariance of each point of interest will collapse identically as a function of the agent's distance from the target. The isotropic measurement covariance model is intentionally adopted as a simplified approximation of sensing performance to isolate the coupling between trajectory distance and covariance evolution. 

The goal of the inspection framework is to drive the estimation covariance of the state $\vec\theta$ toward a desired distribution, and the measurement covariance serves as the coupling between the agent dynamics and the evolution of the covariance over time. 

\section{Approach}
To address this inspection problem, an MPC framework is formulated where a Kalman filter is used to propagate and update the estimates of the points of interest. The Kalman filter equations are additionally enforced as an equality constraint in the optimization to predict how the estimation covariance will evolve due to the computed control inputs. The MPC framework additionally incorporates the measurement model developed above for the Kalman filter, along with the HCW dynamics for the agent motion. This method enables the generation of inspection trajectories that explicitly reduce the estimation covariance of the points of interest to a desired terminal distribution.

\subsection{Kalman Filter and Covariance Dynamics}  The system for the target points of interest defined in Eqs.~\eqref{targetState}-\eqref{targetMeasurement} is linear and coupled to the dynamics of the agent through the measurement covariance in Eq.~\eqref{measurementCovDefinition}. To estimate the points of interest, we choose to apply the linear-discrete Kalman filter \cite{Simon_OptimEstim} with time propagation (\textit{a priori}) estimates 
\begin{align}
    \vec\theta_i^-&=\mathbf A_\theta\vec\theta_{i-1}^+ + \vec w_i, \label{KF_timePropagate1}\\
    \mathbf P_i^- &= \mathbf A_\theta\mathbf P_{i-1}^+\mathbf A_\theta\transpose + \mathbf W, \label{KF_timePropagate}
\end{align}
Kalman gain 
\begin{equation}\label{KF_KGain}
    \mathbf K_i=\mathbf K(\vec x_i,\mathbf P_i^-) = \mathbf P_i^-\mathbf C\transpose(\mathbf C\mathbf P_i^-\mathbf C\transpose+\mathbf V_i)^{-1}, 
\end{equation}
and measurement update (\textit{a posteriori}) estimates 
\begin{align}
    \vec\theta_i^+&=\vec\theta_i^- +\mathbf K_i(\vec y_i-\mathbf C\vec\theta_i^-), \label{KF_measurementUpdate1}\\
    \mathbf P_i^+ &= (\mathbf I-\mathbf K_i\mathbf C)\mathbf P_i^-(\mathbf I-\mathbf K_i\mathbf C)\transpose + \mathbf K_i\mathbf V_i\mathbf K_i\transpose. \label{KF_measurementUpdate}
\end{align}This method of the \textit{a posteriori} update in Eq.~\eqref{KF_measurementUpdate} is the Joseph stabilized update; it ensures that the covariance matrix remains symmetric positive definite in the measurement update provided that the covariance in the time propagation step is positive definite  \cite{Simon_OptimEstim}, which is important for retaining well-defined uncertainty in each point of interest. The covariance equations associated with the Kalman filter above (Eqs.~\eqref{KF_timePropagate}, \eqref{KF_KGain}, and \eqref{KF_measurementUpdate}) are utilized in the MPC framework to predict the evolution due to the computed control inputs.

\subsection{Finite-Horizon Stochastic MPC Problem}
\begin{figure}[htpb]
    \centering
    \includegraphics[width=1\linewidth]{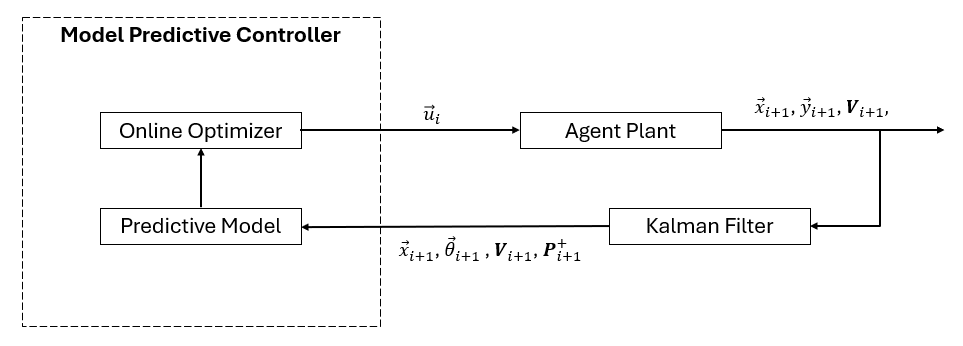}
    \caption{MPC closed-loop diagram. }
    \label{fig:blockDiagram}
\end{figure}

To generate information-aware inspection trajectories, a finite-horizon stochastic MPC framework is employed. The proposed formulation follows the philosophy of explicit dual control (EDC) \cite{Mesbah2018} by incorporating an uncertainty metric directly into the MPC objective, thereby encouraging trajectories that improve future estimation performance without solving the Bellman equation. At each sampling instant, the optimization predicts both the relative motion of the agent and the evolution of the estimation covariance associated with the target points of interest over a finite prediction horizon. Consequently, the optimizer determines a sequence of control inputs that simultaneously reduce control effort and estimation uncertainty while satisfying the spacecraft dynamics and operational constraints. Because the estimation covariance evolves through a trajectory-dependent measurement covariance, predicted information gain directly influences the selected control policy. The MPC controller is implemented in a receding-horizon fashion through the repeated solution of the constrained nonlinear optimization problem shown in Figure~\ref{fig:blockDiagram}. After each optimization, only the first control input is applied to the spacecraft. The system state is then propagated, new measurements are obtained, and the measurement covariance is evaluated. A Kalman filter updates the estimated target features and their associated covariance, which are provided to the next MPC iteration together with the updated spacecraft state. The prediction horizon is shifted forward one time step, and the process is repeated until the final simulation time.

 The resulting finite-horizon optimization problem is formulated as follows: 
\begin{align}
    \min_{\vec\alpha,\vec\beta,\vec\gamma}&~~J_N= \sum_{k=0}^{N-1}\left[\vec\beta_k\transpose \mathbf R\vec\beta_k + S~ \trace{\mathbf\Gamma_k}\right]+S~ \trace{\mathbf\Gamma_N}, \label{problemStatement}\\
    \text{subject~}&\text{to} \nonumber\\
    &\vec\alpha_0=\vec x(t_i), \label{solverState} \\
    &\vec\alpha_{k+1}=\mathbf A_d\vec\alpha_k+\mathbf B_d\vec\beta_k,~~k=0,\dots,N-1, \label{dynamicsConstr} \\
    &||\mathbf H\vec {\alpha}_k|| \geq r_{min},~~k=0,\dots,N,  \label{minDist}\\
    &||\mathbf H\vec {\alpha}_{N}|| \leq r_{max}, \label{maxDist}\\
    &\|\vec\beta_k\|_\infty\leq u_{max},~~k=0,\dots,N-1,  \label{maxThrust} \\
    &\mathbf \Psi_0 = \mathbf V(\vec x(t_i)), \label{solverMeasureCov}\\
    &\mathbf \Psi_{k+1} = (\zeta||\mathbf H\vec {\alpha}_{k+1}||^2 + \varepsilon)\mathbf I_p,~~k=0,\dots,N-1, \label{measureCovProp} \\
    &\mathbf \Gamma_0 =\mathbf P(t_i), \label{solverCov} \\ 
    &\mathbf \Gamma_{k+1} =F(\mathbf\Gamma_{k},\mathbf \Psi_{k+1}(\vec \alpha_{k+1})),~~k=0,\dots,N-1, \label{covPropagation} \\
    &\trace{\mathbf \Gamma_{N}} \leq \Gamma_f, \label{terminalCovConstr} 
\end{align} 
where $\vec\alpha\in\reals^n$, $\vec\beta\in\reals^m$, and $\vec\gamma\in\reals^q$ are the decision variables representing state, control, and covariance as vectors, $\mathbf \Gamma\in\reals^{q\times q}$ represents the covariance in the matrix form, $\mathbf R\in\reals^{m\times m}$ and $S\in\reals_+$ are weighting parameters, $r_{\min}\in\reals_+$, $r_{max}\in\reals_+$, and $u_{max}\in\reals_+$ define distance and control constraints, $\mathbf\Psi\in\reals^{p\times p}$ represents measurement covariance, and $\Gamma_f\in\reals_+$ defines the terminal covariance constraint. The cost function in Eq.~\eqref{problemStatement} contains a running cost on control effort and estimation covariance along with a terminal cost on covariance. The relative weighting between control $\vec\beta_k$ and covariance $\mathbf\Gamma_k$ determines the tradeoff in maneuvering effort and information acquisition. By having a running cost on covariance, the optimizer is incentivized to find a trajectory that improves the information gained along the entire trajectory, not just at the end. The constraints in Eqs.~\eqref{solverState}-\eqref{terminalCovConstr} encode agent dynamics, safety, covariance propagation, and the final covariance goal for the inspection mission.

To reduce the number of decision variables, the formulation exploits a storage vector $\vec \gamma_k$ as the decision variables corresponding to unique terms of the covariance matrix $\mathbf\Gamma_k$. Covariance matrices are inherently symmetric and positive semi-definite, and are in fact positive definite assuming that all channels, or sources of measurements, are linearly independent \cite{BendatPiersol_randomData}. Thus, only the diagonal and upper-triangular off-diagonal elements must be stored explicitly. 

For a target with eight points of interest, where each point of interest contains a $3\times 3$ covariance matrix corresponding to position uncertainty in the LVLH frame, the covariance matrix $\mathbf\Gamma_k$ is structured as a block diagonal matrix
\begin{equation}
\mathbf{\Gamma}_k =
\begin{bmatrix}
\mathbf{\Gamma}_{1,k} & & \\
& \ddots & \\
& & \mathbf{\Gamma}_{8,k}
\end{bmatrix},
\end{equation} where each block is symmetric and given by \begin{equation}
\mathbf{\Gamma}_{j,k} =
\begin{bmatrix}
\gamma^{(j)}_{11,k} & \gamma^{(j)}_{12,k} & \gamma^{(j)}_{13,k} \\
\gamma^{(j)}_{21,k} & \gamma^{(j)}_{22,k} & \gamma^{(j)}_{23,k} \\
\gamma^{(j)}_{31,k} & \gamma^{(j)}_{23,k} & \gamma^{(j)}_{33,k}
\end{bmatrix},
\end{equation} for $j=1,\dots,8$ points of interest. The corresponding storage vector therefore contains only the unique entries from each covariance block:
\begin{equation}
\vec{\gamma}_k =
\begin{bmatrix}
\gamma^{(1)}_{11,k} &
\gamma^{(1)}_{22,k} &
\gamma^{(1)}_{33,k} &
\gamma^{(1)}_{12,k} &
\gamma^{(1)}_{13,k} &
\gamma^{(1)}_{23,k} &
\cdots &
\gamma^{(8)}_{23,k}
\end{bmatrix}^{T}.
\end{equation}
By assuming that the initial covariance, process and measurement covariances, state matrix of the target states, and the output matrix of the measurement model are all diagonal matrices, the Kalman filter as in Eqs.~\eqref{KF_timePropagate1}-\eqref{KF_measurementUpdate} produces diagonal Kalman gains and preserves the diagonal structure of the propagated covariance. Therefore, to further reduce the number of decision variables associated with the covariances of the points of interest and reduce computation for the optimization, the storage vector is constructed as
\begin{equation}
\vec{\gamma}_k =
\begin{bmatrix}
\gamma^{(1)}_{11,k} &
\gamma^{(1)}_{22,k} &
\gamma^{(1)}_{33,k} &
\gamma^{(2)}_{11,k} &
\gamma^{(2)}_{22,k} &
\gamma^{(2)}_{33,k} &
\cdots &
\gamma^{(8)}_{33,k}
\end{bmatrix}^{T},
\end{equation} where the entries of $\vec\gamma_k$ are only the diagonal entries of each covariance block.

For this formulation, the current time index corresponding to real time is $i\in\mathbb W$, while the time steps over the finite horizon $N$, for one iteration of MPC, are $k=0,\dots,N$. This notation clarifies that plant parameters $\vec x_i,~\vec u_i,~\mathbf V(\vec x_i),~\mathbf{P}(t_i)$ are separate from the solver parameters $\vec\alpha_k,~\vec\beta_k,~\mathbf\Psi_k,~\mathbf \Gamma_k$. As such, when constructing the constraints of the MPC framework, the solver is initialized with the current state, measurement covariance, and estimation covariance $\vec x_i,~\mathbf V(\vec x_i),~\mathbf{P}(t_i)$ as the initial solver states, measurement covariance, and estimation covariance $\vec\alpha_0,~\mathbf\Psi_0,~\mathbf \Gamma_0$ in Eqs.~\eqref{solverState}, \eqref{solverMeasureCov}, \eqref{solverCov}. These initialization equations are the bridge between the agent plant parameters and the MPC solver parameters.

The problem is constrained by the HCW dynamics of the agent in Eq.~\eqref{dynamicsConstr}, where $(\mathbf A_d,\mathbf B_d)$ are the discrete-time system matrices from Eqs.~\eqref{discreteStateMatrix} and \eqref{discreteControlMatrix}. The relative distance between the agent and the target is constrained by $r_{min}$ in Eq.~\eqref{minDist} to avoid collision with the target. The relative distance is constrained by $r_{max}$ in Eq.~\eqref{maxDist} as a terminal constraint of each horizon to ensure the trajectory can return to a desired inspection range between the agent and target, as the cost function does not contain a state regulation term to maintain stable trajectories. The maximum absolute value of the control input $\vec\beta_k\in\reals^m$ is bounded by $u_{max}$ in Eq. \eqref{maxThrust} to maintain physical limits set by the thrusters on the agent.  

The measurement covariance is updated following the definition in Eq.~\eqref{measureCovProp} to represent how measurement quality changes with distance. The estimation covariance matrix is propagated via a nonlinear function $F(\mathbf\Gamma_{k},\mathbf \Psi_{k+1}(\vec \alpha_{k+1}))$ in Eq.~\eqref{covPropagation} that encompasses the Kalman filter \textit{a priori} and \textit{a posteriori} steps as in Eqs.~\eqref{KF_timePropagate1}-\eqref{KF_measurementUpdate}. A terminal constraint is placed on the trace of the final covariance of the horizon to ensure that the solver finds a trajectory that reduces the overall covariance on the state of the target points of interest in Eq.~\eqref{terminalCovConstr}.

Once an iteration of MPC is complete, the initial control input $\vec \beta_0$ from the computed control over the prediction horizon is assigned as the control input to the plant as $\vec u_i$. At the next time instances $i+1,~i+2,\dots$, the agent is propagated along its trajectory, and the MPC scheme is computed iteratively. To initialize each MPC iteration, the solver is warm-started using a shifted initial guess of the prediction horizon, where the initial decision variables are trimmed, and a final guess is computed to append at the end of the horizon.

\section{Results} 

\subsection{Example Inspection Trajectory} The above problem is applied to a specific inspection scenario, where Table 1 summarizes the parameters chosen for the following results. The nonlinear optimization problem was solved using the sequential quadratic program (SQP) implemented through MATLAB’s \texttt{fmincon} solver \cite{MATLABfmincon}.
\begin{table}[htbp]
	\fontsize{10}{10}\selectfont
    \caption{Simulation Parameters}
   \label{tab:paramChoices}
        \centering 
   \begin{tabular}{c | r | r  | r} 
      \hline 
      Parameter & Symbol & Value & Units\\
      \hline 
        Max iterations & - & 200 &- \\
        Max function evaluations & - & 1E+4& -\\
        Step tolerance & - & 1E-12& -\\
        Constraint tolerance & - & 1E-6& -\\
        Discretization step & $\Delta t$ & 60 & s\\
        MPC prediction horizon length & \textit N & 8 & time steps\\
        Simulation time & $t_f$ & 5.553E+3& s\\
        Mean motion of target & \textit n & 0.0011 & rad/s\\
        Orbital period of target & \textit T & 5.553E+3& s\\
        Target cube side length & \textit d & 1& m\\
        Mass of agent & \textit m & 50 & kg\\
        Initial state of agent & $\vec x(t_0)$ & [0, 1, 0, 5.657E-4, 0, 1.1314E-4]$\transpose$& km, km/s\\
        Initial estimation covariance & $\mathbf P(t_0)$ &  $\mathbf I_{24}$ & km$^2$\\
        Measurement covariance scaling term & $\zeta$ & 1E-5& -\\
        Min measurement variance & $\varepsilon$ & 1E-8& km$^2$\\
        Process covariance & $\mathbf{W}$ & 1E-6 *$\mathbf I_{24}$ & km$^2$\\
        Control cost weight & $\mathbf R$ & $\mathbf I_3$& -\\
        Covariance cost weight & $S$ & 10 & -\\
        Min distance constraint & $r_{min}$ & 5E-2& km\\
        Terminal max distance constraint & $r_{max}$ & 1.5 & km\\
        Max control constraint & $u_{max}$ & 1 & N\\
        Terminal covariance constraint & $\Gamma_f$ & 5E-6& km$^2$\\
      \hline
   \end{tabular}
\end{table} The target's orbit was chosen to be a circular low Earth orbit (LEO) with a radius of 6778 km, where the radius defines the mean motion $n$ and the orbital period $T$. The initial state of the agent was chosen to be in a natural motion circumnavigation (NMC), which provided a stable relative orbit derived from the HCW equations \cite{NMTs}. As an initial guess of the first horizon in the MPC problem, this initial condition was propagated assuming zero control. Due to the stability of the NMC with respect to the position of the target, this provided a feasible initial guess for the SQP solver. From the initial guess of the trajectory, the measurement covariance was computed using Eq.~\eqref{measurementCovDefinition} and the estimation covariance was propagated using the Kalman filter from Eqs.~\eqref{KF_timePropagate} and \eqref{KF_measurementUpdate}. The agent was propagated for one orbital period of the target, $T$. 

\begin{figure}[htpb]
    \centering
    \includegraphics[width=0.55\linewidth]{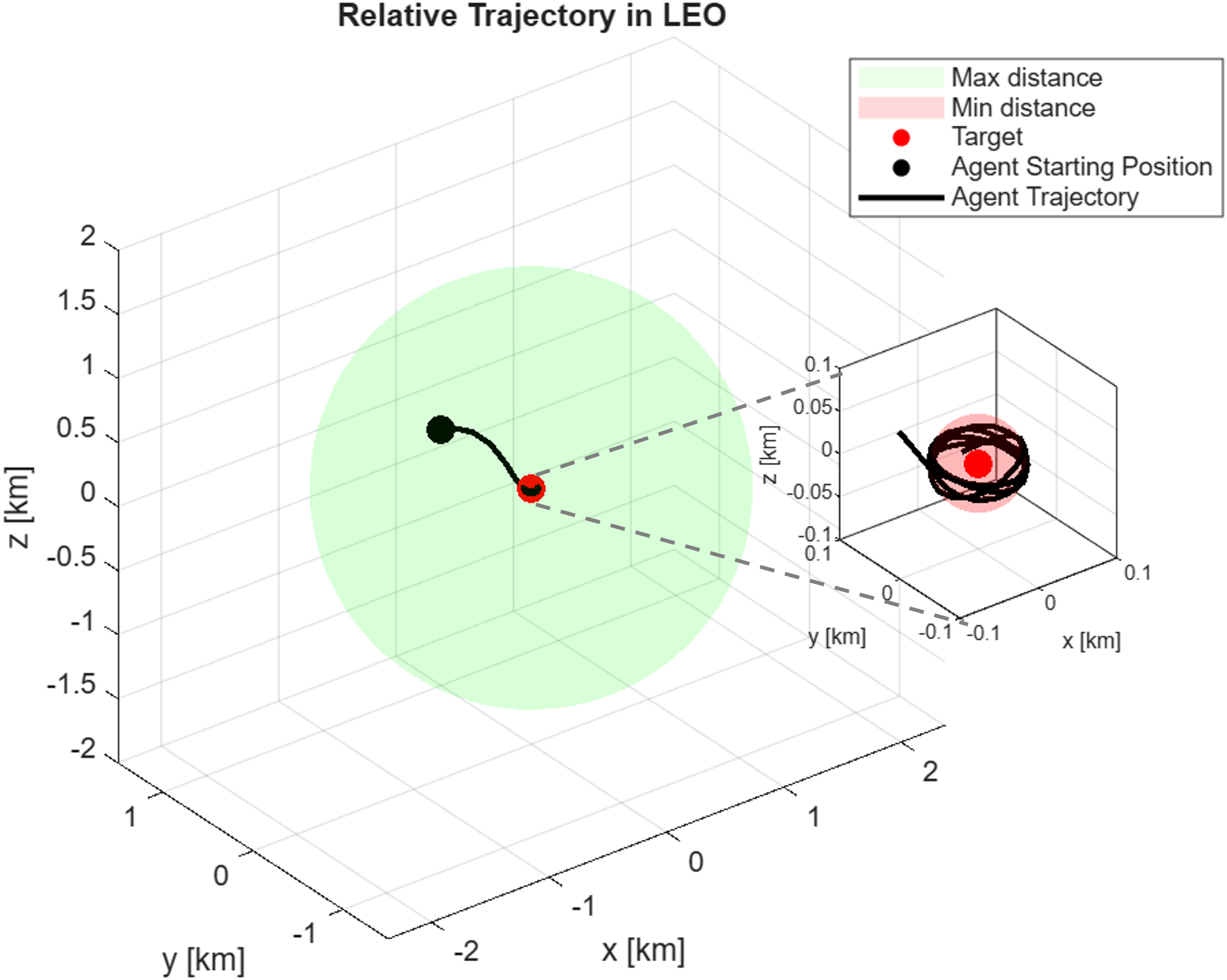}
    \caption{Agent trajectory in the LVLH frame with the minimum and maximum distance constraint boundaries shown.}
    \label{fig:trajectory}
\end{figure}

\begin{figure}[htpb]
    \centering
    \includegraphics[width=0.65\linewidth]{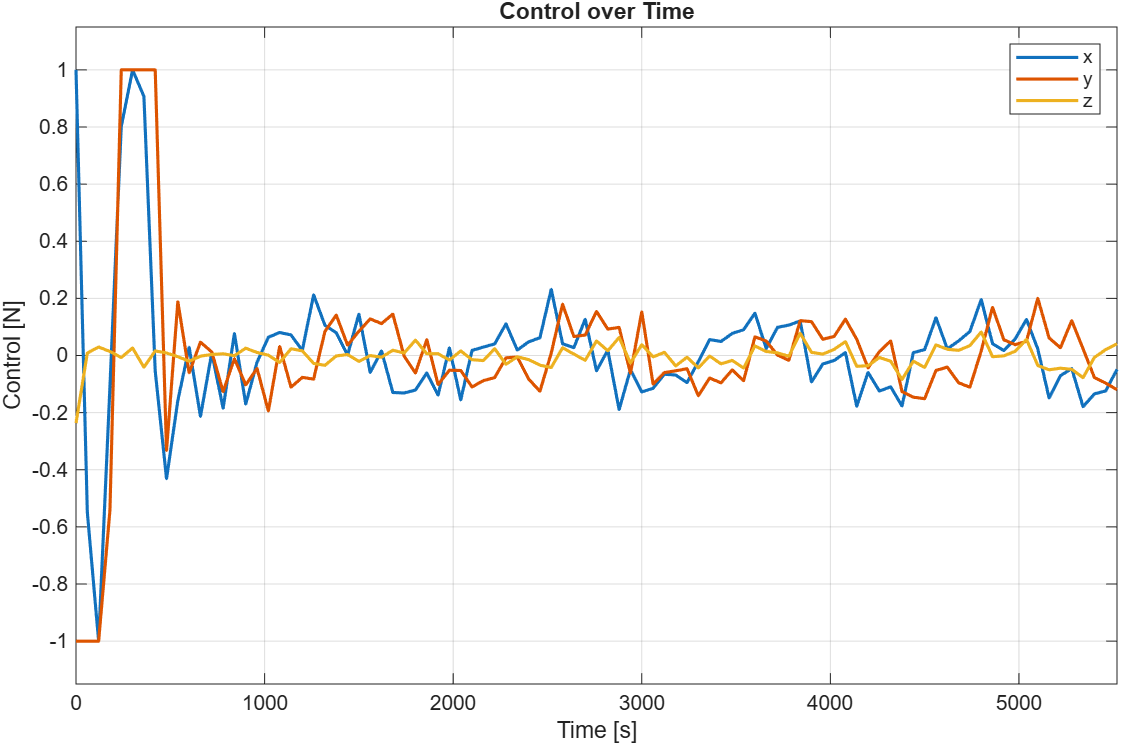}
    \caption{Control input of the agent as seen in the LVLH frame of the target. The control input is constrained to 1 N along each LVLH axis $x,y,z$.}
    \label{fig:control}
\end{figure}

In the trajectory, the agent begins by approaching the target;  once the agent reaches the minimum distance constraint for collision avoidance, the agent orbits about the target following the constraint boundary (Figure \ref{fig:trajectory}).
This behavior is due to the construction of the measurement covariance; since the cost penalizes nonzero estimation covariance, the incentive is to reduce the distance between the agent and the target to minimize measurement covariance, as it reduces the estimation covariance. However, the distance is constrained by the minimum distance constraint of 0.05 km, so the trajectory cannot continue converging to the target to minimize the measurement covariance. The control input (Figure \ref{fig:control}) shows that the control remains feasible with respect to the imposed thrust constraints on each control axis. During approximately the first 480 seconds of the maneuver, the control approaches the maximum thrust constraint of 1 $\mathrm{N}$ as the optimizer drives the agent toward the target to reduce estimation covariance. After the trajectory reaches the minimum-distance constraint boundary, the control effort decreases substantially as the agent transitions to maintaining a bounded relative orbit near the constraint surface.

\begin{figure}[htbp]
\centering
\includegraphics[width=0.65\textwidth]{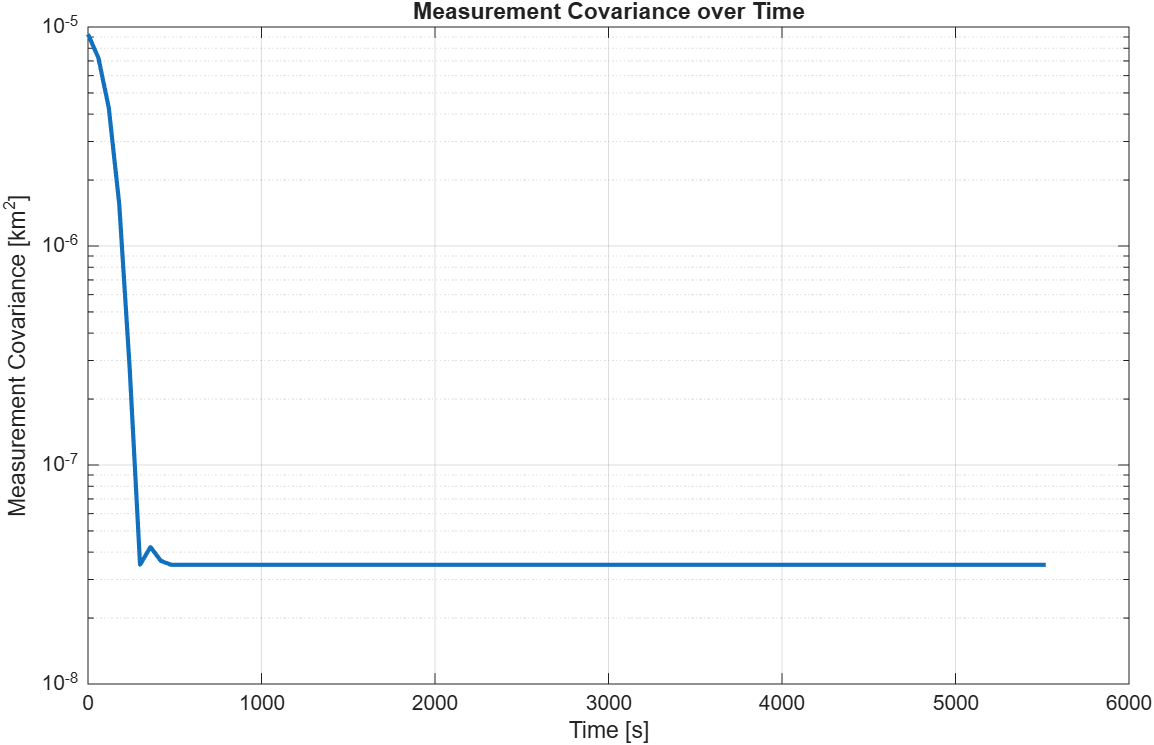}
\caption{Measurement covariance over time. The measurement covariance is bounded below by the minimum distance constraint between the agent and the target.}
\label{fig:measurementCovariance}
\end{figure}

\begin{figure}[htpb]
    \centering
    \subfigure[Trace of the estimation covariance of each point of interest over time. The covariance of each point of interest was initialized to 1 km$^2$ along each axis, resulting in an initial trace of 3 km$^2$ per point. The terminal constraint is shown as a dotted line, and is the threshold for the sum of the individual traces of each point.]{
    \includegraphics[width=0.65\textwidth]{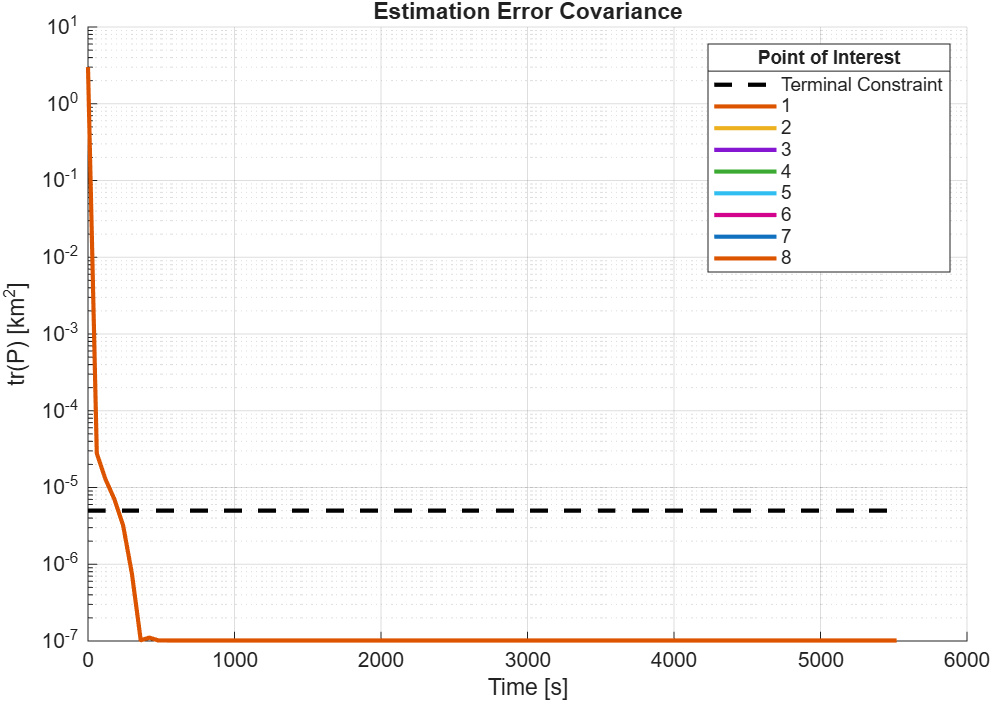}
    }
    \subfigure[Covariance ellipsoids for the estimated points of interest over time overlaid on the target cube geometry. Note the change in axis scale between the first and second plots, as the estimator is initialized with a covariance of 1 km$^2$ along each axis.]{
    \includegraphics[width=1\textwidth]{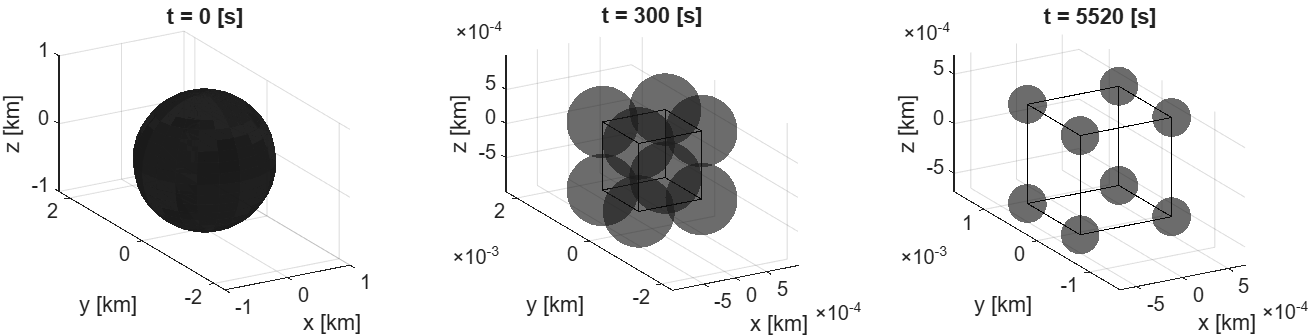} }
    \caption{Estimation covariance.}
    \label{fig:estimationCovariance}
\end{figure}

 The measurement covariance for each point of interest on the target is seen in Figure \ref{fig:measurementCovariance}. At approximately $t=480$ s, the agent reaches the minimum-distance constraint boundary, after which the measurement covariance approaches a near-constant value of 3.5E-8 $\mathrm{km}^2$. This lower bound arises from the distance constraint and the dependence of distance in the measurement covariance formulation. The estimation covariance associated with each point of interest decreases accordingly (Figure \ref{fig:estimationCovariance}). This behavior results from the isotropic structure of the measurement covariance together with the identity output matrix assumed in the measurement model, which causes the information gain to depend primarily on the relative distance between the agent and the target rather than viewing geometry. 

\begin{figure}[htpb]
    \centering
    \includegraphics[width=0.65\linewidth]{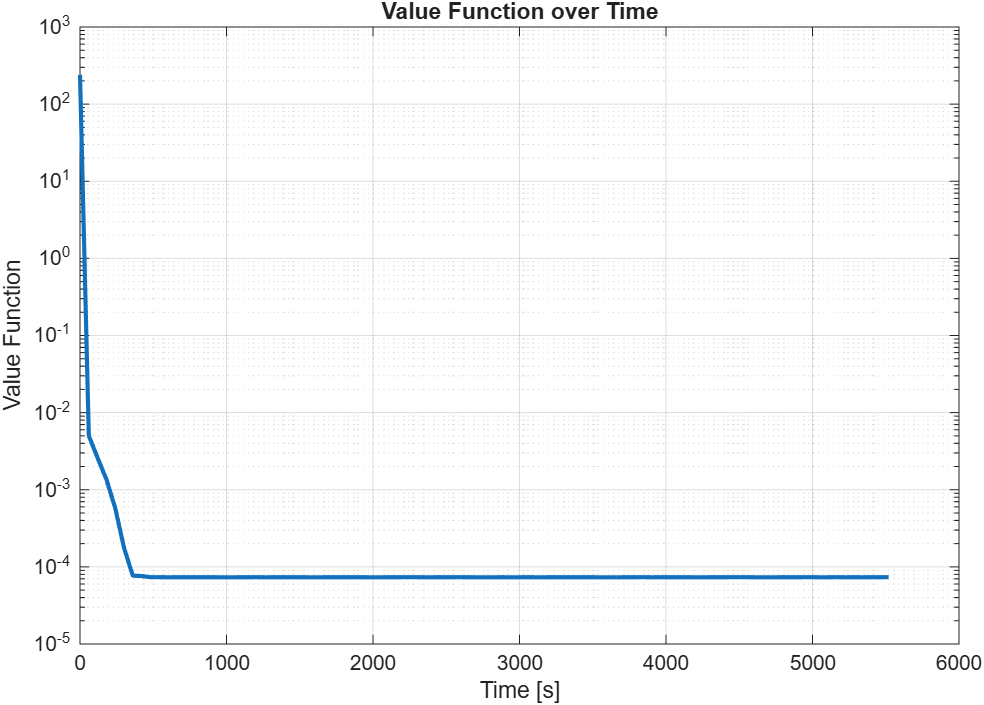}
    \caption{Value function of the MPC solution.}
    \label{fig:valueFunction}
\end{figure}

The value function, corresponding to the minimized MPC cost at each prediction horizon, illustrates the optimization behavior of the MPC framework (Figure \ref{fig:valueFunction}). Once the agent reaches the minimum-distance constraint boundary, the value function approaches a near steady-state value. This behavior occurs because both dependencies of the cost function, the control effort and the trace of the estimation covariance, remain approximately constant after the trajectory settles near the constraint boundary. 

These results demonstrate the effectiveness of the proposed approach in reducing estimation uncertainty while maintaining a feasible trajectory. The value function decreases monotonically over time and approaches a near steady-state regime once the agent reaches the minimum-distance constraint boundary. At this stage, the control effort and estimation covariance remain approximately constant due to the balance between information collection, process noise, and the enforced distance constraint. Because the process covariance remains nonzero, the estimation covariance does not asymptotically converge to zero, but instead approaches a bounded neighborhood determined by the process and measurement covariances. This behavior suggests that practical stability concepts may apply to this system under the proposed MPC framework. 

\subsection{Initial Condition Feasibility Mesh Analysis} A second study was conducted to vary the semi-minor axis length $b$ and starting in-plane angle $\phi$ of the agent's initial NMC relative to the target to investigate how feasibility and value function vary with initial condition. For each initial condition, the parameters chosen in Table 1 were held constant, except for the choice in $\vec x(t_0)$, as a function of $b\in\{0.05,10\}$ $\mathrm{km}$ and $\phi\in\{0^{\circ},360^{\circ}\}$ \cite{NMTs}. 

\begin{figure}[htpb]
    \centering
    \includegraphics[width=0.65\linewidth]{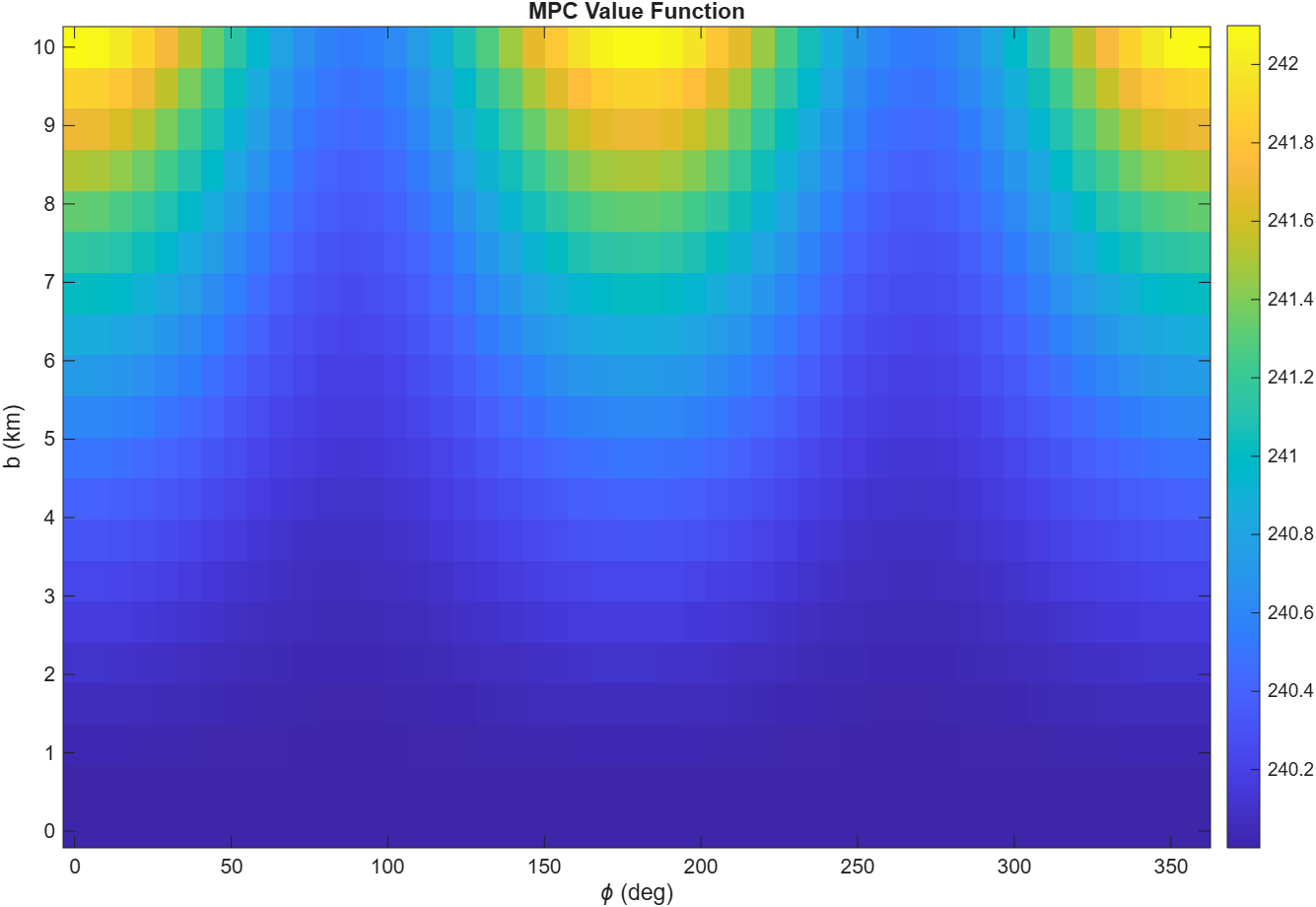}
    \caption{Value function of the first prediction horizon as a function of the agent's initial condition.}
    \label{fig:valueFunctMesh}
\end{figure}

The value function represents the optimal cumulative cost over the first horizon for each choice of initial condition $b,\phi$ of the starting NMC for the agent (Figure \ref{fig:valueFunctMesh}). Particularly, for a fixed $\phi$, decreasing $b$ generally reduces the value function. However, for a fixed $b$, the value function is reduced as $\phi$ approaches values of 90$^{\circ}$ and 270$^{\circ}$. From the formulation of the NMC as an initial condition, when the starting angle $\phi$ is at a value of 90$^{\circ}$ or 270$^{\circ}$, the distance between the agent and the target is at its smallest value, $b$. When the starting angle is at a value of 0$^{\circ}$ or 180$^{\circ}$, the distance between the agent and the target is at its largest, $2b$.  Therefore, the general trend is that the cost over the first horizon decreases as the initial distance of the agent from the target decreases. 

\begin{figure}[htpb]
    \centering
    \includegraphics[width=0.6\linewidth]{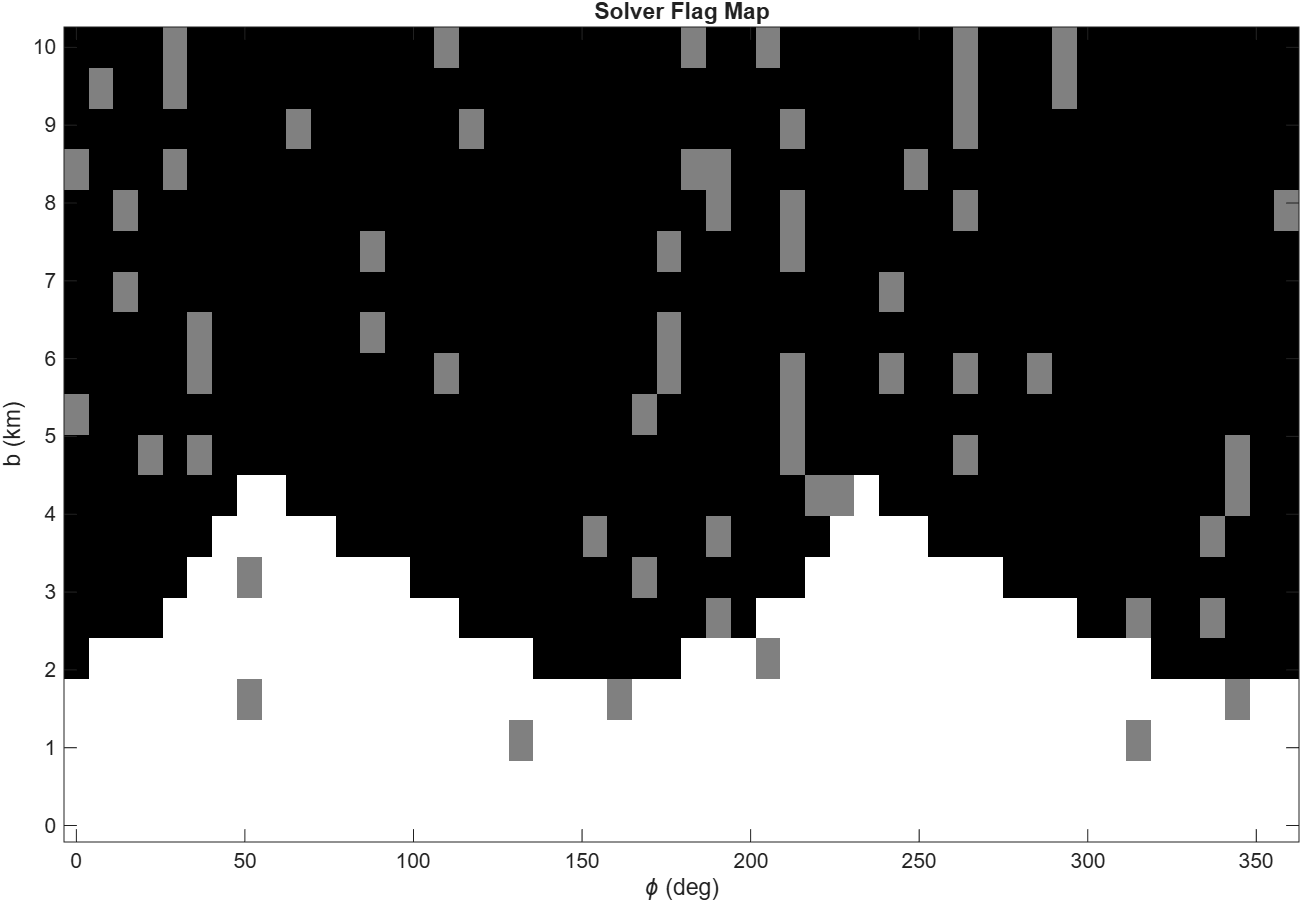}
    \caption{\texttt{fmincon} flag output. Black indicates infeasible solutions that converged, and white indicates feasible solutions that converged. Gray indicates suboptimal solutions that terminated prior to convergence due to the SQP iteration limit. Feasibility of the gray points depends on the constraint feasibility shown in Figure \ref{fig:constraintFeasibility}.}
    \label{fig:feasibilityMesh}
\end{figure}

However, not all initial conditions result in a feasible solution (Figure \ref{fig:feasibilityMesh}). In the solver \texttt{fmincon}, the solution has a diagnostic flag that indicates convergence and feasibility \cite{MATLABfmincon}. Particularly, if a solution contains \texttt{flag=1}, the solution converged to a feasible point. If it contains \texttt{flag=-2}, then the solution converged to an infeasible point. If it contains $\texttt{flag=0}$, the solution did not converge due to solver iteration limits; the solution may be infeasible or feasible. If the solution did not converge, but is feasible, then the solution is suboptimal. These feasibility results indicate that not all solutions of the value function are valid, so Figure \ref{fig:valueFunctMesh} must be interpreted carefully; only feasible, optimal solutions result in a valid value function. 

\begin{figure}[htpb]
    \centering
    
    \subfigure[Constraint feasibility for control magnitude. Distances around 4 km or greater tend to lead to an infeasible control magnitude.]{
    \includegraphics[width=0.6\textwidth]{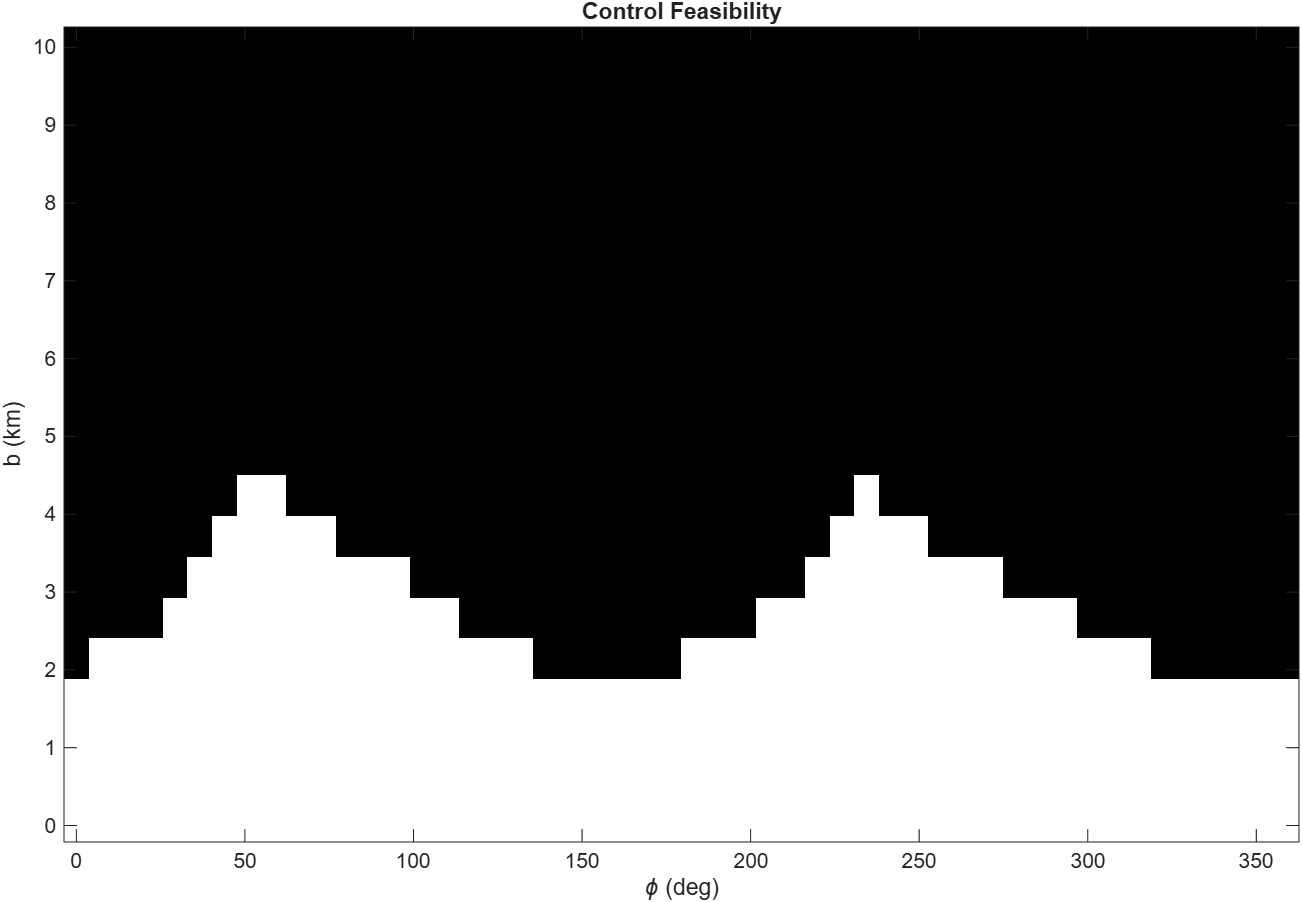} }
    
    \subfigure[Constraint feasibility for terminal maximum distance. Generally, the constraint is feasible except for initial conditions with the largest distances from the target, as $\phi=0,180,360^\circ$ correspond to when the agent begins at a distance of $2b$.]{
    \includegraphics[width=0.6\textwidth]{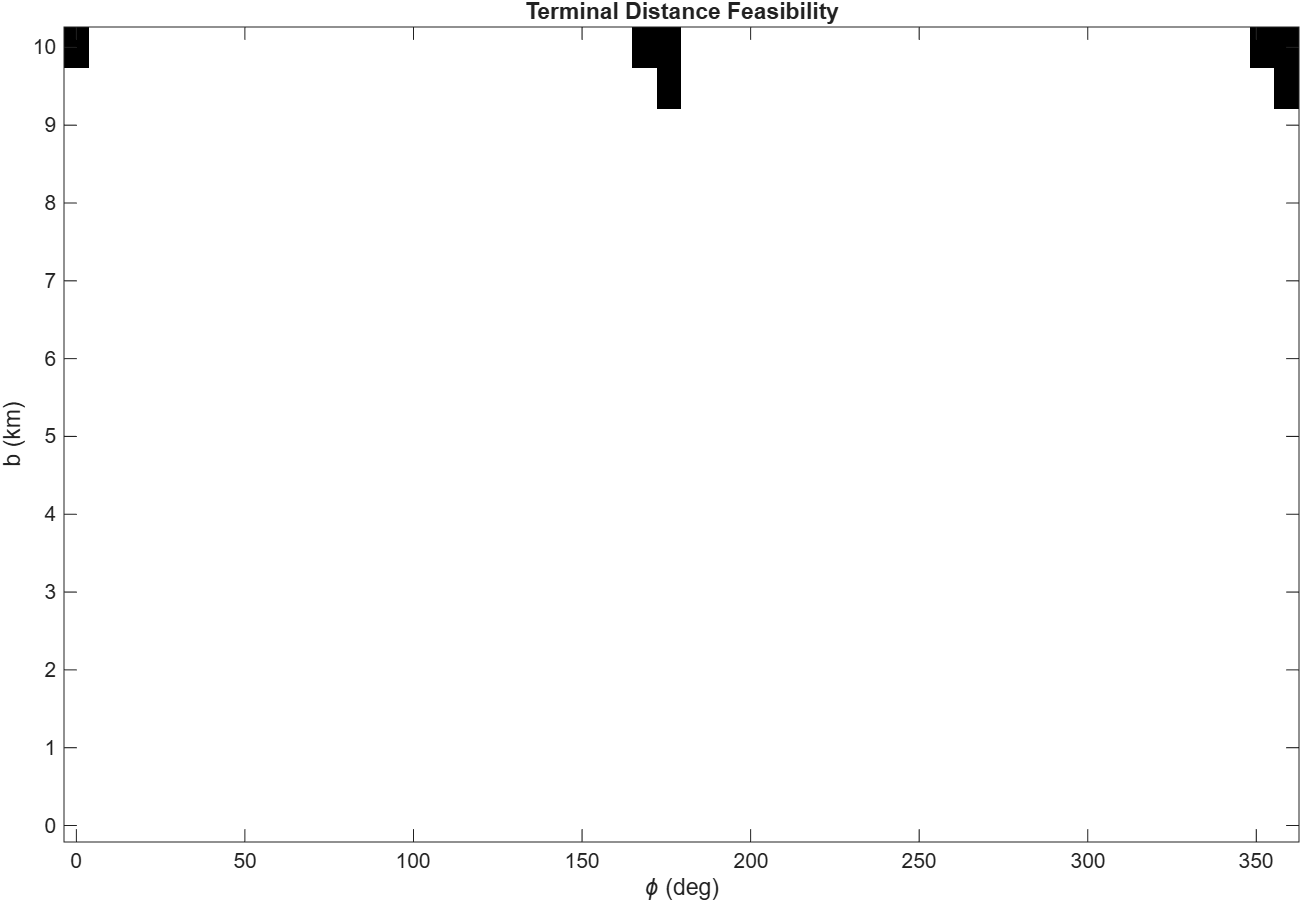}
    }

    \caption{Constraint feasibility for inequality constraints. Black indicates infeasible solutions, and white indicates feasible solutions. The minimum distance and terminal covariance constraints remain feasible.}
    \label{fig:constraintFeasibility}
\end{figure}

Plotting the individual inequality constraints reveals whether the gray points are feasible (Figure \ref{fig:constraintFeasibility}). All initial conditions were feasible in the terminal covariance constraint or the minimum distance constraint. However, many initial conditions with distances of about 4 km or greater tended to exceed actuation limits, and some initial conditions near the maximum initial starting distance, which corresponds to starting along the semi-major axis of the NMC where $b=10~\mathrm{km}$ and $\phi=0^{\circ},180^{\circ}$, failed to close in the distance between the agent and the target to achieve the terminal maximum distance. This infeasibility suggests that the control constraint, which must hold at each time step, and the terminal maximum distance constraint, which only holds at the end of the horizon, are too strict for specific agent initial conditions in a finite horizon $N$. So, careful design of the agent initial condition is required for feasibility when spacecraft thruster capabilities are limited. 

    \label{fig:ConstraintFeasibility}

\section{Conclusion}
 This work presents an information-aware MPC framework for autonomous spacecraft inspection using concepts from dual control and covariance steering. By incorporating Kalman filter covariance propagation into a constrained MPC formulation, the proposed approach explicitly couples estimation performance with the control policy through a geometry-dependent measurement covariance model. As a result, the generated inspection trajectories not only satisfy relative motion and thrust constraints, but also actively reduce uncertainty in the estimated features of a target spacecraft. 
 
 Simulation results demonstrate the feasibility of the proposed framework and illustrate how sensing geometry influences the resulting inspection trajectory. Particularly, the optimizer drives the agent toward a relative orbit that improves the estimation while remaining within the feasible region defined by distance constraints. The initial condition mesh analysis further demonstrates that feasibility and value function behavior are coupled to the initial position of the agent with respect to the target, with infeasible regions arising primarily from actuation and terminal distance limitations within a finite prediction horizon.  
 
 The current formulation adopts an isotropic measurement covariance model with direct measurements of each point of interest. Ongoing work will extend the measurement model through a geometry-dependent output matrix incorporating intermittent measurements due to sensor field-of-view constraints and target self-occlusion. These effects will produce estimation dynamics in which covariance evolution depends explicitly on visibility conditions and measurement availability through the agent trajectory. In addition, future work will investigate theoretical stability properties of the system to determine whether practical stability guarantees can be established for the MPC framework.


\bibliographystyle{AAS_publication}  
\bibliography{referencesClees}  


\end{document}